\documentclass{elsart}
\usepackage{amsmath}
\usepackage{amsfonts}
\usepackage{amssymb}
\usepackage{graphicx}
\newcommand{\singlefig}{.75\textwidth}
\newcommand{\doublefig}{\textwidth}
\begin{document}
\date{19 March 2002}
\begin{frontmatter}

\title{Nonlinear charge transport mechanism in periodic and disordered DNA}

\author[Berlin]{D. Hennig\thanksref{cor}},
\author[Sevilla]{J.F.R. Archilla},
\author[Berlin]{J. Agarwal},
\thanks[cor]{Corresponding author.
             E-mail:hennigd@physik.fu-berlin.de}

\address[Sevilla]{Departamento de F\'{\i}sica Aplicada I.
 Universidad de Sevilla.\\
    Avda. Reina Mercedes, s/n. 41012-Sevilla, Spain}

\address[Berlin]{Freie Universit\"{a}t Berlin, Fachbereich Physik,
Institut f\"{u}r Theoretische Physik.\\ Arnimallee 14, 14195 Berlin,
Germany}

\journal{Physica D}

\begin{abstract}

\noindent We study a model for polaron-like charge transport
mechanism along DNA molecules with emphasis on the impact of
parametrical and structural
disorder. Our model Hamiltonian takes into
account the coupling of the charge carrier to two
different kind of modes representing fluctuating twist motions of
the base pairs and H-bond distortions within the double helix
structure of $\lambda-$DNA. Localized stationary states are
constructed with the help of a nonlinear map approach for a periodic
double helix and in the presence of
 intrinsic static parametrical and/or structural disorder reflecting the
impact of ambient solvent coordinates.
It is demonstrated that charge transport is mediated by
 moving polarons respectively breather compounds
 carrying not only the charge but
causing also
local temporal deformations of the helix structure through the
traveling torsion and bond breather components illustrating  the
interplay of structure and function in biomolecules.

\end{abstract}


\end{frontmatter}


\section{Introduction}

Electronic transport through DNA is crucial for its biological
functions such as the repair mechanism after radiation damage and
biosynthesis.  As proposed in \cite{Eley} electron transport
through DNA proceeds along a one-dimensional pathway constituted
by the overlap between $\pi-$orbitals in neighboring base pairs.

With the perspective to possible applications in molecular
electronics biomaterials are supposed to play an important role
because they may offer the ultimate way to miniaturization and their
attributed
electrical transport properties are expected to be markedly
different from those of traditional macroscopic conductors \cite{Mirkin}.
In this context enormous effort has been directed towards investigations of
conductivity of DNA viewed as a promising conduit for charge
transport on nanoscales \cite{Ratner}. So far the findings
concerning conductivity of DNA have been controversial. There are
the results of  recent measurements indicating that  DNA behaves
as a well conducting one-dimensional molecular wire
\cite{Hall}-\cite{Tran}.
In contrast it was reported that DNA is insulating
\cite{Braun} and for short oligomers built up from
 base pairs of the same type semiconductivity was observed
\cite{Porath}. This qualitative discrepancy is shared also by
theoretical findings so that it remains unclear whether DNA is
conductive or not. For example a theoretical investigation based
on first principle calculations suggests the absence of
dc-conductivity in DNA \cite{Pablo}. On the other hand there stand
the studies using various assumptions for modeling the DNA
structure which  focus on different aspects such as the influence
of aperiodicity, temperature driven fluctuations and  aggregation
effects on the energetic control of charge migration which
have all come to the conclusion that DNA is a conductor
\cite{Jortner}-\cite{Hjort}.

The detailed understanding of the conduction mechanism of DNA
remains still an unsolved task.  Different attempts to model  the
charge transport of DNA were based on transport via coherent
tunneling \cite{Eley}, classical diffusion under the
conditions of temperature-driven fluctuations \cite{Bruinsma},
incoherent phonon-assisted hopping \cite{Ly},\cite{Jortner1},
variable range hopping between localized states \cite{Yu} and charge
carriages mediated by polarons \cite{Conwell},\cite{Ly1} and
solitons \cite{Hermon}.

In the present work we use for the theoretical description of
charge transport in DNA a nonlinear approach facilitating the
concept of polaron respectively breather solutions. The structure
of the bent double helix of $\lambda-$DNA is modeled by a sterical
network of oscillators in the frame of the base pair picture
\cite{PB},\cite{Yakhu},\cite{Peyrard} taking into account
deformations of the hydrogen bonds within a base pair and twist
motions between adjacent base pairs. The electron motion is described
by a tight-binding system. It is assumed that the electron motion is
predominantly influenced by vibrational modes of the double helix.
The nonlinear interaction between the
electron and the vibrational modes causes the formation of polarons
respectively electron-vibron breathers
supporting charge localization and transport.

The aim of the current investigation is to elucidate the role
played by static parametrical disorder and complex structural
effects due to irregular deviations from the ordered helical shape
of the double helix (reflected in random arrangements of the
positions of the bases) with regard to the charge transport in
DNA. There arises at least one interesting question, namely: Is
 long-lived and stable charge transport in DNA achievable
at all under the impact of disorder\,?
To tackle this problem in our paper we exploit analytical and numerical methods from
 the theory of (nonlinear) dynamical systems in combination with the
theory of (linear) disordered systems.  The paper is organized as follows: In  the first
section
we describe  the model for the charge transport along the steric
structure of the bent DNA double helix. The second section is
concerned with the construction of polaron respectively static
electron-vibron breather solutions of the nonlinear lattice system
when  parametrical and structural disorder effects are taken into account.
In section \ref{section:moving} we discuss the mobility properties of
the polarons  respectively electron-vibron breathers and their
transport efficiency in dependence on the  degree of
disorder contained in
the on-site energies and/or in structural arrangements of the helix. In a subsequent section

we
invoke the Floquet analysis to explain also the mobility of the polarons
and breathers in terms of their spectral properties.
Finally, we present a summary and discussion of the results.

\section{Model Hamiltonian for charge transfer along DNA}

\begin{figure}[h]
  \begin{center}
    \includegraphics[width=\singlefig]{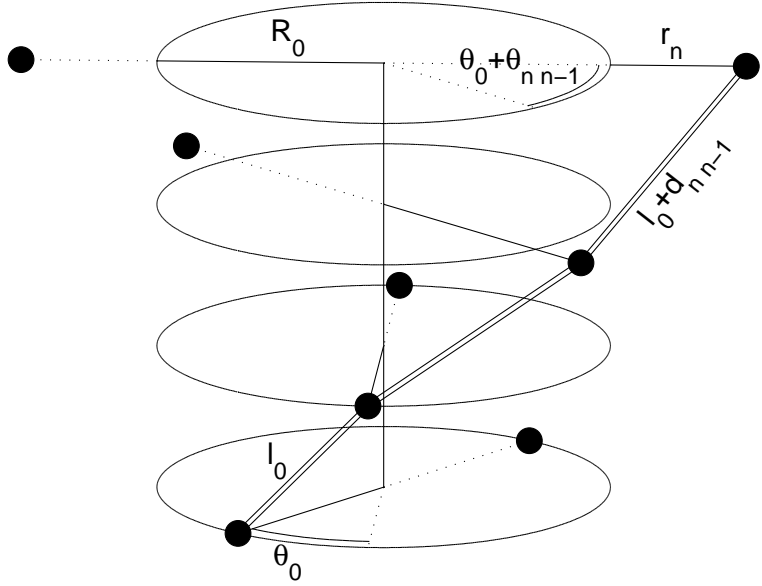}
  \caption{Sketch of the helicoidal structure of
the DNA model, the bases being represented by bullets. Geometrical
parameters $R_0$, $\theta_0$, $l_0$ and the radial and angular
variables $r_n$ and $\theta_{n\,n-1}$, respectively are indicated.
}
  \label{fig:sketch}
 \end{center}
\end{figure}

DNA chains are quite flexible and exhibit  structural fluctuations
influencing the transport properties. Our model is designed to cover
the basic features of the DNA double helix structure needed for
a proper description of the charge transport dynamics under the
impact of fluctuating base motions. Despite the fact that the
charge transport proceeds across a one-dimensional channel of base
stacks spanning along the sugar-phosphate backbone so that DNA can
effectively be viewed as a one-dimensional molecular wire
\cite{Stryer} some significant details of the three-dimensional
DNA structure have to be incorporated in the model. For the
physical description of DNA molecules their detailed chemical
structure can be neglected. They can be simply considered as bent
double-stranded systems with an adequate number of parameters
 sufficient for modeling the considered physical process.
In the frame of simple DNA models such as the base pair picture
the bases have been treated as single nondeformable entities. The
helicoidal structure of DNA is then conveniently described in a
cylindrical reference system where each base pair possesses two
degrees of freedom, namely a radial variable measuring the
transversal displacements of the base pair (that is deformations
of the H-bond) and its angle with a reference axis in a plane
perpendicular to the helix backbone  which defines the twist of
the helix \cite{Peyrard}. In Fig.~\ref{fig:sketch} 
we depict schematically the
structure of the DNA model.

With relevance to charge migration the most influential fluctuating
motions of DNA originate from tranverse vibrations of the bases relative
to each other within each base plane, i.e. the base pair distance
stretchings/compres\-sions and the torsional variations of the
helicoidal twist dominating longitudinal (acoustic strand) phonons \cite{Bruinsma}
because the sugar-phosphate chains are rather stretch proof
\cite{Stryer}. Therefore it is justified to discard the impact of
longitudinal vibrations and the considered motion is restricted to the
base pair planes \cite{Ye}. Furthermore, the angular twist and radial
vibrational motions evolve independently on two different time scales
and can be regarded as decoupled degrees of freedom in the harmonic
treatment of the normal mode vibrations of DNA \cite{Cocco}.

The Hamiltonian for the electron migration along a strand in DNA
consists of three parts
\begin{equation}
H=H_{el}+H_{rad}+H_{twist}\,,
\end{equation}
with $H_{el}$ representing  the part corresponding to the particle
charge transport over the base pairs,
 $H_{rad}$ describes  the  dynamics of the
H-bond vibrations and $H_{twist}$ models the dynamics of the
relative twist angle between two consecutive base pairs. The
electronic part is given by a tight-binding system
\begin{equation}
H_{el}=\sum_{n} \,E_n\,|c_{n}|^2
-V_{n\,n-1}\,\left(\,c_{n}^{*} c_{n-1}+c_n c_{n-1}^{*}\,\right)
\,.\label{eq:Hel}
\end{equation}
The  index  $n$ denotes the site of the $n-$th base on one of the
two strands  and $c_n$ determines the probability that the
electron occupies this site.  $E_n$ is the local electronic energy
and $V_{nn-1}$ stands for the transfer matrix element which is responsible
for the transport of the electron along the stacked base pairs.

The vibronic part $H_{rad}$ takes into account radial
displacements of the base units from their equilibrium positions
within the base pair plane connected with deformations of the
hydrogen bonds linking two bases. The corresponding part of the
Hamiltonian treating the bond vibrations classically and
harmonically is given by
\begin{equation}
H_{rad}= \,\sum_{n}\,\left[\,\frac{
\left(p^{\,r}_{n}\right)^2}{2\,M_n}\,+\,\frac{M_n\,\Omega_{r}^2\,r_n^2}{2}\,\right]\,.
\end{equation}
with  the momenta $(p_n^{\,r})^2$ conjugate to the radial
coordinates $r_{n}$ measuring the deviation from the equilibrium
length of a hydrogen bond along the line bridging two bases of a
base pair, $\Omega_{r}$ is the frequency and $M_n$ is the reduced
mass of a base pair.

 The twist motion part reads as
\begin{equation}
H_{twist}=\,\sum_{n}\,
 \large[\,\frac{ \left(p^{\,\theta}_{n\,n-1}\right)^2}{2\,J_n}\,
+\,\frac{J_n\,\Omega_{\theta}^2\,\theta_{n\,n-1}^2}{2}\,\large]\,,\label{eq:Htwist}
\end{equation}
where $\theta_{n\,n-1}$ is the relative angle between
 two adjacent
base pairs quantifying displacements from its equilibrium
twist angle $\theta_0$ and $J_n$ is the reduced moment of inertia.

We emphasize that in the
context of charge transport in DNA  the nonlinear dynamics of
large amplitude structural transitions are not of interest because charge
transport processes are thought to be mainly facilitated by soft fluctuational
vibrational modes allowing the harmonic approximation of their
dynamics.

The interaction between the electronic variable and the structure
variables $r_{n}$ and $\theta_{n\,n-1}$ stems from the parameter
dependence of the electronic parameters $E_n$ and $V_{n\,n-1}$.
The  diagonal term expressing the most efficient coupling is of
the form
\begin{equation}
E_{n}=E_{n}^{0}+k\,r_{n}\,,
\end{equation}
and takes into account  the modulation of the
on-site electronic energy $E_{n}^{0}$ by the radial vibrations of
the base pairs. In turn the actual charge occupation has its
impact on the local radial distortion of the helix. As the
transfer matrix elements $V_{n\,n-1}$ are concerned we assume that
they depend on the three-dimensional distance between two consecutive bases along a
strand in the following fashion
\begin{equation}
V_{n\,n-1}=V_0\,(1-\alpha\,d_{n\,n-1})\,.
\end{equation}
The quantity $\alpha$ regulates how strong $V_{n\,n-1}$ is
influenced by the distance and the latter is determined by
\begin{eqnarray}
d_{n\,n-1}&=&\left\{a^2+(R_0+r_n)^2+(R_0+r_{n-1})^2\right.\nonumber\\
&-&\left.2(R_0+r_n)(R_0+r_{n-1})\,\cos(\theta_0+\theta_{n\,n-1})
\right\}^{1/2}-l_0\,, \label{eq:dist}
\end{eqnarray}
with
\begin{equation}
l_0=\sqrt{a^2+4R_0^2\,\sin^2(\theta_0/2)}\,.
\end{equation}
Expanding the expression (\ref{eq:dist}) up to first order around the
equilibrium positions gives
\begin{equation}
d_{n\,n-1}\simeq \frac{R_0}{l_0}\,\left[\, (\,1-\cos
\theta_0\,)\,(r_n+r_{n-1})\,+\,\sin \theta_0\,
R_0\,\theta_{n\,n-1}\,\right]\,.\label{eq:distance}
\end{equation}

Realistic parameters for DNA molecules are given by
 \cite{Peyrard},\cite{Stryer}:
$a=3.4\AA$, $R_0 \thickapprox 10\AA$, $\theta_0=36^\circ$,
$J=4.982\times10^{-45}\,kg\,m^2$, $\Omega_{\theta}=[0.526
-0.744]\,\times10^{12}\,s^{-1}$,
$\Omega_{r}=6.252\,
\times10^{12}\,s^{-1}$, $V_0\simeq 0.1\,eV$ and
$M=4.982\times10^{-25}kg$.

 We scale the time according to
$t\rightarrow \Omega_{r}\,t$ and introduce the
dimensionless quantities:
\begin{eqnarray}
\tilde{r}_{n}&=&\sqrt{\frac{M \Omega_{r}^{2}}{V_{0}}}\, r_{n}\,,
\qquad \tilde{k}_{n}=\frac{k_n}{\sqrt{M\Omega_{r}^{2}V_0}}\,,
\qquad \tilde{E_n^0}=\frac{E_n^0}{V_0}\\
\tilde{\Omega}&=&\frac{\Omega_{\theta}}{\Omega_{r}}\,, \,\,\,\,
\tilde{V}=\frac{V_0}{J\,\Omega_{r}^{2}}\,,\,\,\,\,
\tilde{\alpha}=\sqrt{\frac{V_0}{M\,\Omega_{r}^2}}\,\alpha\,,
\,\,\,\, \tilde{R}_{0}=\sqrt{\frac{M\,\Omega_{r}^2}{V_0}}\,R_{0}
\,.
\end{eqnarray}
Subsequently we omit the tildes.

With the use of the expression (\ref{eq:distance}) the equations of
motion derived from the Hamiltonian
(\ref{eq:Hel})-(\ref{eq:Htwist}) read as
\begin{eqnarray}
i\,\tau\dot{c}_{n}&=&(E_n^0\,+k\,r_n)\,c_n\nonumber\\
&-&(1-\alpha\,d_{n+1,n})\,c_{n+1} -(1-\alpha\,d_{n\,n-1})\,c_{n-1}
\label{eq:dotc}\\ \ddot{r}_{n}&=&-r_n-k\,|c_n|^2\,-\,\alpha\,
\frac{R_0}{l_0}\,(1-\cos \theta_0)\,\nonumber\\ &\times&
\left\{\,[c_{n+1}^*c_{n}+c_{n+1}c_{n}^*]+[c_{n}^*c_{n-1}+c_{n}c_{n-1}^*]\,\right\}\label{eq:dotr}\\
\ddot{\theta}_{n\,n-1}&=&-\Omega^2\,\theta_{n\,n-1}\,-\,
\alpha\,V\,\frac{R_0^2}{l_0}\, \sin
\theta_0\,[c_{n}^*\,c_{n-1}\,+\,c_{n}\,c_{n-1}^*]\,,
\label{eq:dottheta}
\end{eqnarray}
and the quantity $\tau=\hbar\,\Omega_{r}/V_0$ appearing in
Eq.~(\ref{eq:dotc}) determines the time scale separation between the
fast electron motion and the slow bond vibrations. We remark that in the
limit case of $\alpha=0$ and constant $E_n^0=E_0$ the set of coupled
equations represents the Holstein system widely used in studies of
polaron dynamics in one-dimensional lattices \cite{Holstein}.
Furthermore in the linear limit case emerging for $\alpha=k=0$
 the Anderson model is obtained for random $E_n^0$ \cite{Anderson},\cite{Mott}.

Regarding the dynamical energy exchange between the various
degrees of freedom we state that due to the large differences
between the electronic and vibronic frequencies we are able to
prove rigorously that the energy exchange is suppressed on time
scales growing exponentially with the ratio of the frequencies.
Corresponding bounds on the energy exchange suppression are
derived with the help of Nekhoroshev-like arguments
\cite{Bambusi},\cite{Hennigmath}. Furthermore applying the concept of the
anti-integrable limit \cite{MacKay} we are able to prove the
existence of breather solutions for the coupled system
(\ref{eq:dotc})-(\ref{eq:dottheta}). The mathematical details are
delegated to a forthcoming article \cite{proof}.

The values of the scaled parameters are given by $\tau=0.2589$,
$\Omega^2=[\,0.709-1.417\,]\,\times 10^{-2}$, $V=0.0823$,
$R_0=34.862$ and $l_0=24.590$. In our model study we treat the
electron-mode coupling strengths $k$ and $\alpha$ for which no
reliable data are available as adjustable parameters.
In the forthcoming investigations we adopt the values of these
free parameters such that not too strong deformations of the helix
result which is essential for our used harmonic treatment of the
dynamics of the structural coordinates.

\section{Localized polaron-like states}\label{section:polaron}

Being interested in the simulation of a nonlinear charge transport
mechanism in DNA we embark as a first step on the construction of
localized stationary solutions of the coupled system
(\ref{eq:dotc})-(\ref{eq:dottheta}). Exploiting the fact that the
adiabaticity parameter $\tau$ is small, i.e. the fast charge
transport along the base stacks and the slow bond vibrations and
the even slower rotational twist motions evolve on distinct time
scales, the inertia in Eqs. (\ref{eq:dotr}) and
(\ref{eq:dottheta}) are negligible. Therefore, one can solve in
the adiabatic limit the resulting  static equations which results
in the instantaneous displacements
\begin{eqnarray}
r_{n}&=& -k\,|c_n|^2\,-\,\alpha\, \frac{R_0}{l_0}\,(1-\cos
\theta_0)\,\nonumber\\ &\times&
\left\{\,[c_{n+1}^*c_{n}+c_{n+1}c_{n}^*]
+[c_{n}^*c_{n-1}+c_{n}c_{n-1}^*]\,\right\}\,, \label{eq:rinst}\\
\theta_{n\,n-1}&=&-\frac{\alpha\,V}{\Omega^2}\,
\frac{R_0^2}{l_0}\,\sin \theta_0\, [c_n^*c_{n-1}+c_nc_{n-1}^*]
\label{eq:thetainst}\,.
\end{eqnarray}
Upon insertion of (\ref{eq:rinst}) and (\ref{eq:thetainst}) into
Eq.~(\ref{eq:dotc}) one obtains a nonlinear discrete Schr\"{o}dinger
equation for the electronic amplitude
\begin{eqnarray}
i\,\tau\,\dot{c}_{n}&=&\left[E_n^0-k^2\,|c_n|^2-k\,\alpha\,
\frac{R_0}{l_0}\,(1-\cos \theta_0)\right. \nonumber\\
&\times&\left.\left\{\,[c_{n+1}^*c_{n}+c_{n+1}c_n^*]
+[c_n^*c_{n-1}+c_n c_{n-1}^*]\,\right\}\,\right]\,c_n \nonumber\\
&-&(\,1+b_{n+1\,n}\,)\,c_{n+1}\, -\,(\,1+b_{n\,n-1}\,)\,c_{n-1}
\label{eq:DNLS}\,,
\end{eqnarray}
with
\begin{eqnarray}
b_{n\,n-1}&=& \alpha\,\left(\frac{R_0}{l_0}\right)^2\,\left\{\,
(1-\cos \theta_0)\, \left[\,k\,\frac{l_0}{R_0}\,
[|c_n|^2+|c_{n-1}|^2]\right.\right.\nonumber\\ &+&
\left.\alpha\,(1-\cos \theta_0)\,
\left(\,[c_{n+1}^*c_{n}+c_{n+1}c_n^*]\right.\right.\nonumber\\
&+&\left.\left.2\,[c_n^*c_{n-1}+c_n c_{n-1}^*]
+[c_{n-1}^*c_{n-2}+c_{n-1} c_{n-2}^*]\,\right)\right]\nonumber\\
&+&\left. \frac{\alpha\,V}{\Omega^2}\,R_{0}^2\,\sin^2\theta_0\,
[c_n^*c_{n-1}+c_n c_{n}c_{n-1}^*]\right\}
\end{eqnarray}

In order to search for stationary localized solutions (polaron states)
we substitute $c_{n}=\Phi_{n}\,\exp[-i\,Et/\tau]$ in Eq.~(\ref{eq:DNLS})
and obtain the difference system
\begin{eqnarray}
E\,\Phi_{n}&=&\left[E_n^0-k^2\,|\Phi_n|^2-k\,\alpha\,
\frac{R_0}{l_0}\,(1-\cos \theta_0)\right. \nonumber\\
&\times&\left.\left\{\,[\Phi_{n+1}^*\Phi_{n}+\Phi_{n+1}\Phi_n^*]
+[\Phi_n^*\Phi_{n-1}+\Phi_n \Phi_{n-1}^*]\,\right\}\,\right]
\,\Phi_n \nonumber\\ &-&(\,1+B_{n+1\,n}\,)\,\Phi_{n+1}\,
-\,(\,1+B_{n\,n-1}\,)\,\Phi_{n-1} \,, \label{eq:statthree}
\end{eqnarray}
where $B_{n\,n-1}$ is given by
\begin{eqnarray}
B_{n\,n-1}&=& \alpha\,\left(\frac{R_0}{l_0}\right)^2\,\left\{\,
(1-\cos\theta_0)\, \left[ k\,\frac{l_0}{R_0}\,
[|\Phi_n|^2+|\Phi_{n-1}|^2]\right.\right.\nonumber\\ &+&
\left.\alpha\,(1-\cos \theta_0)\,
\left(\,[\Phi_{n+1}^*\Phi_{n}+\Phi_{n+1}\Phi_n^*]
\right.\right.\nonumber\\
&+&\left.\left.2\,[\Phi_n^*\Phi_{n-1}+\Phi_n \Phi_{n-1}^*]
+[\Phi_{n-1}^*\Phi_{n-2}+\Phi_{n-1} \Phi_{n-2}^*]\,\right) \right]
\nonumber\\ &+&\left.
\frac{\alpha\,V}{\Omega^2}\,R_{0}^2\,\sin^2\theta_0\,
[\Phi_n^*\Phi_{n-1}+\Phi_n \Phi_{n-1}^*]\right\}
\end{eqnarray}

The ground state of the system (\ref{eq:statthree}) is computed
with the numerical map method outlined in \cite{Kalosakas},\cite{Nick}.

\begin{figure}[h]
  \begin{center}
    \includegraphics[width=\singlefig]{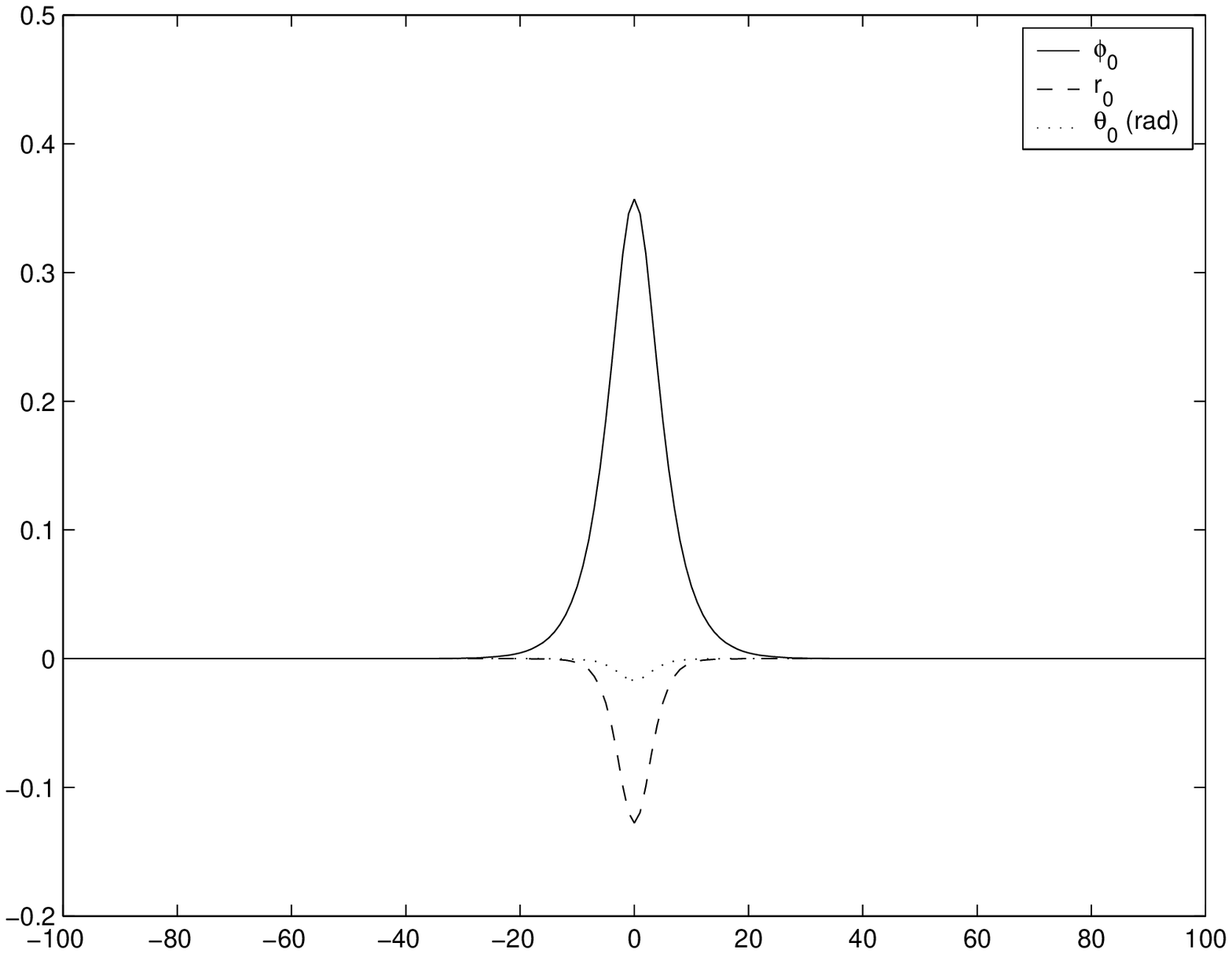}
  \caption{Profiles of the stationary polaron
states for the ordered (periodic) case.
\newline (a) The electronic
amplitudes $|\Phi_{n}|^2$.
 \newline (b) The static radial displacements $r_n$.
 \newline (c) The static angular twists $\theta_n$.}
  \label{fig:profile}
\end{center}
\end{figure}

In Fig.~\ref{fig:profile} 
we depict typical  patterns of
polaron states. First we discuss the ordered (periodic) case
arising e.g. for synthetically produced DNA molecules consisting
of a single type of base pairs (e.g. poly(G)-poly(C) DNA polymers)
surrounded by vacuum which provides realistic ideal conditions for
the consideration of DNA conductivity in experimental and
theoretical studies \cite{Porath}.
 The
electronic contribution is shown in Fig.~\ref{fig:profile}~(a).
 In the ordered case the electronic wave function  is
 localized  at
the central lattice site and the amplitudes decay monotonically
and exponentially with growing distance from this central site. We
remark that  the spatial extension of the polaron decreases with
increased coupling strength $k$ and there is a smooth transition
from large to small polarons. Analogously, enlarging the
off-diagonal coupling strength $\alpha$ leads also to enhanced
degree of localization. The size of the polarons plays a crucial
role for their mobility in the sense that  for large up to medium
polarons coherent motion can be activated whereas small polarons
are immobile due to their pinning to the discrete lattice
\cite{Peyrard1},\cite{Flach}.
The associated patterns of the static radial and angular
displacements of the bases are
 shown in Fig.~\ref{fig:profile}~(b) and (c),
 respectively. Like
 the electronic wave function the radial and angular displacements
  are exponentially localized at the central lattice site (base
  pair). Notice that due to the overall minus sign of the
  radial and angular excitation patterns the H-bridges get compressed
  while the helix experiences a local unwinding around
the occupation peak of the localized electron.

\begin{figure}
  \begin{center}
    \includegraphics[angle=90,height=0.27\textheight]{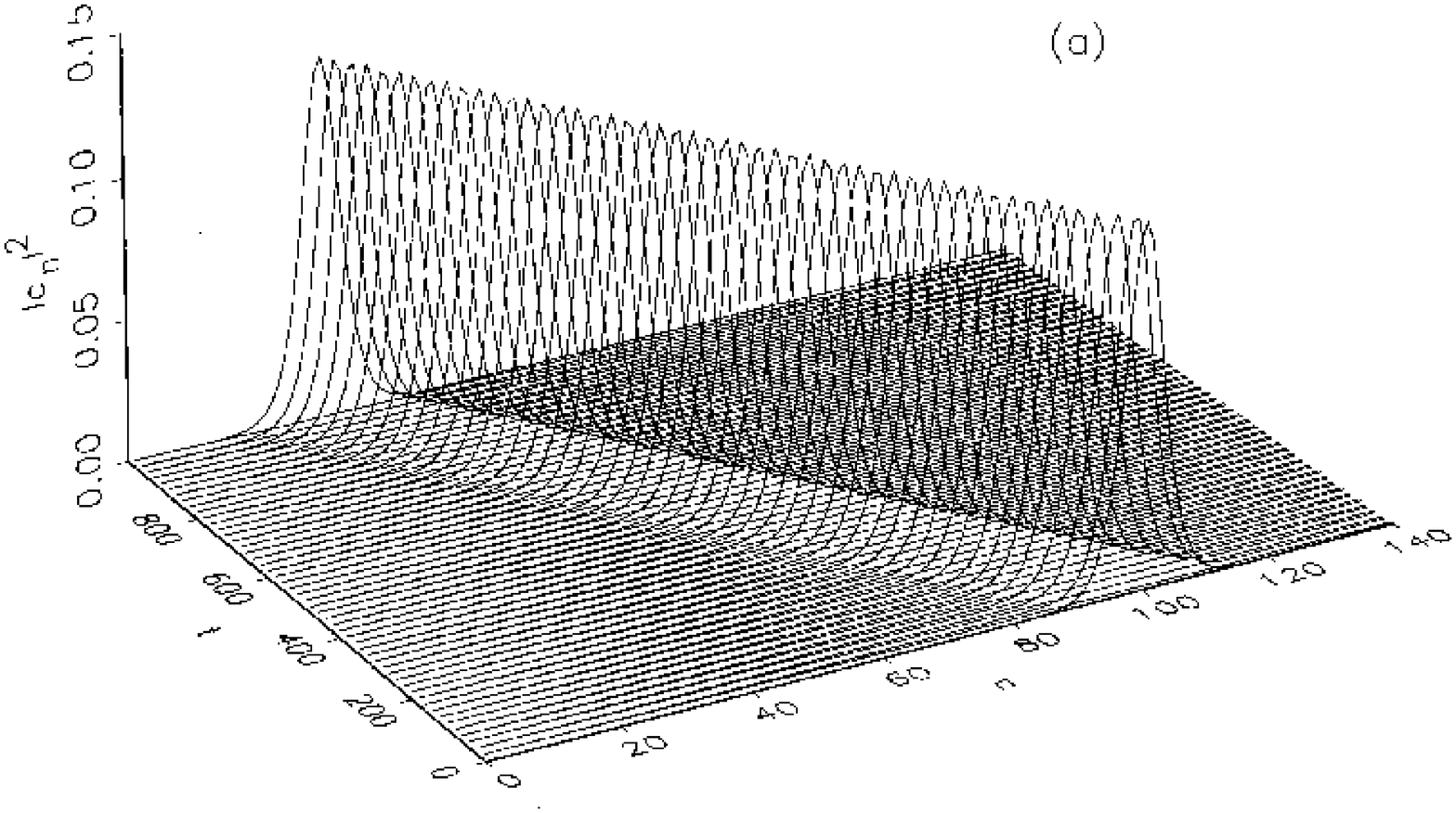}
    \includegraphics[angle=90,height=0.27\textheight]{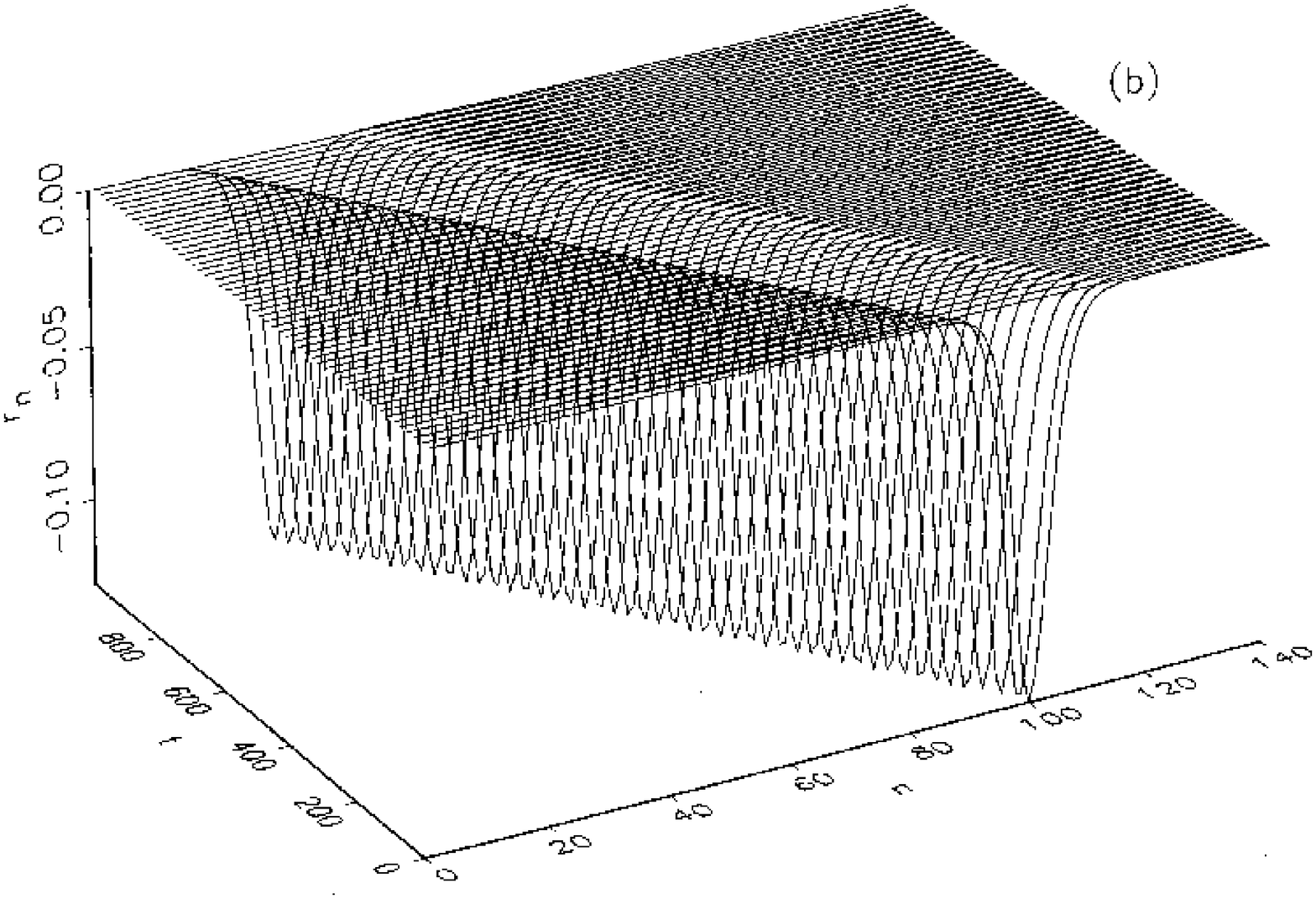}
    \includegraphics[angle=90,height=0.27\textheight]{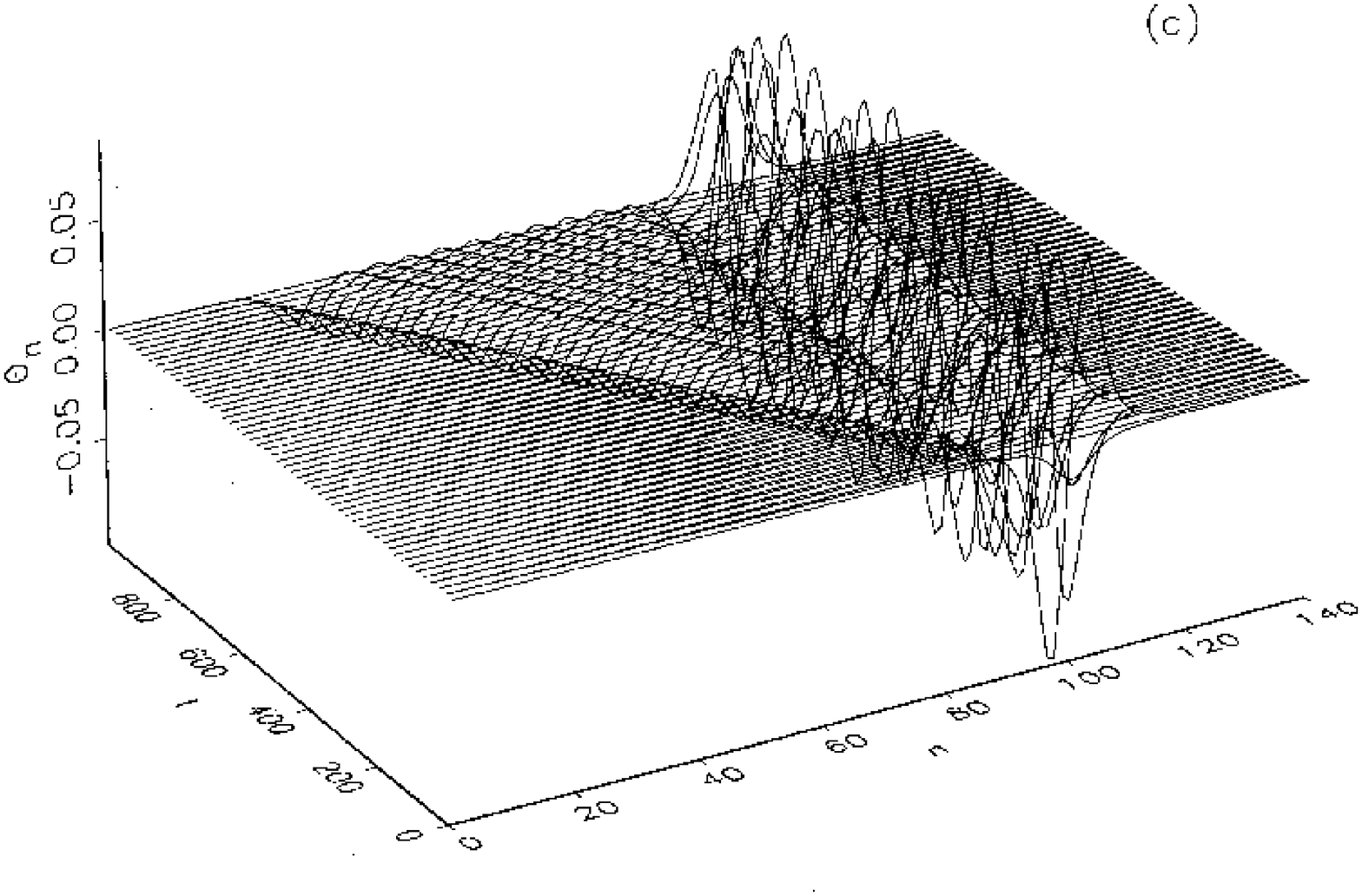}
  \caption{Breather motion along the DNA in the
ordered case. Motion is initiated by an initial kick of the
velocities $\{\dot{r}_{n}\}$ directed along the corresponding
pinning mode component. The kick amplitude is $\lambda_r=0.25$.
\newline (a) Uniformly moving electronic breather.
\newline (b) The associated moving breather in the radial
displacements.
\newline (c) The angular breather contribution consisting of a
small amplitude breather pinned at the starting site and a
relatively large amplitude moving breather component propagating
in alliance with the electron and radial breathers. }
  \label{fig:fig3}
\end{center}
\end{figure}

For a more realistic study we take static diagonal disorder in the
on-site electronic energy $E_n^0$ into account. The random values caused
for example by the inhomogeneous broadening of the sites of distinct ion
pairs with different energies were simulated by random potentials
$[-\Delta E, \Delta E]$ with mean value $\bar{E}=0$ and different mean
standard deviations $\Delta E$. (Note that any $E_{0}c_n$ term on the
r.h.s. of Eq.~(\ref{eq:dotc}) with constant $E_0$ can be eliminated by a
gauge transformation $c_{n}\rightarrow \exp(-iE_{0}t/\tau)c_{n}$.)

Generally the localized excitation patterns do not change
qualitatively when diagonal disorder is taken into account.
Compared with the ordered case one merely observes a shift of the
excitation peak away from the central lattice site depending on
the respective realization of the random potential. However, since
in the presence of diagonal disorder the translational invariance
of the lattice system is broken the localized amplitude pattern is
no longer reflection symmetric with respect to a distinguished
lattice site as opposed to the corresponding symmetric localized
state of the ordered system. Moreover we find that the combined
effect of the two localization mechanisms, viz. nonlinear polaron
and linear Anderson-mode formation respectively, leads to
enhancement of the degree of localization compared to the case
when only one of the two mechanisms acts.

Furthermore, real DNA molecules exhibit random structural
imperfections of their
 double helix caused e.g. by the random base sequence, the deforming
  impact of
the chemical surroundings when DNA  gets buffeted by water
molecules and/or the varying hydrophobic potential of the base
pair interactions depending on the ambient aqueous solvent that
may leave the involved helix structures in irregularly distorted
shapes. Accordingly, the structural disorder is incorporated in
our model by randomly distributed equilibrium positions of the
bases so that the resulting irregular helical DNA matrix deviates
from the perfectly regular helix structure.

To be precise, we consider randomly distributed torsional
coordinates $\theta_{n\,n-1}$ and  radii $r_n$. Since the
positions (actually, the distances) govern the value of the
corresponding transfer matrix elements $V_{n\,n-1}=V_0\,(1-\alpha
d _{n\,n-1})$ nondiagonal disorder is induced in the latter at the
same time. Random transfer integrals along the helix were
used to represent random base pair sequences for which the charge transport
along DNA  was discussed recently in a model relying on localized
electron states \cite{Ye}.

\section{Charge transport mediated by mobile
polarons and breathers}\label{section:moving}

In this section we study the charge transport supported by mobile
polarons respectively breathers propagating along DNA. To activate
polaron motion an established method is provided by proper
stimulation of localized internal normal modes of certain shape
and frequency which are also called pinning modes \cite{Chen}. The
identification  of possibly existing pinning modes is related to the linear stability
analysis of the polarons. In order to investigate the stability of
the polarons with respect to linear perturbations we
 derive the linearized system for which we impose
small perturbations on the stationary polaron solution
$(\Phi_{n}$, $r_{n}^{(0)},
\theta_{n}^{(0)})$ in the form:
\begin{eqnarray}
c_{n}(t)&=&\left[\,\Phi_{n}+ \delta
c_{n}(t)\,\right]\,\exp(-i\,E t/\tau) \label{eq:cperturb}\\
r_{n}(t)&=&r_{n}^{(0)}+ \delta
r_{n}(t)\\ \theta_{n}(t)&=&\theta_{n}^{(0)}+
\delta \theta_{n}(t)\,,\label{eq:thetaperturb}
\end{eqnarray}
where $\delta c_{n}$, $\delta r_{n}$ and
$\delta \theta_{n}$ are small time-dependent
 variables.
Substitution of (\ref{eq:cperturb})-(\ref{eq:thetaperturb}) into
Eqs. (\ref{eq:dotc})-(\ref{eq:dottheta}) and retaining only the
terms linear in the perturbations results in the system of tangent
equations. The stability analysis is then performed using Floquet
theory.

To this end we integrate the linear system of the tangent
equations over one period $T=2\pi\,\tau/E$ yielding the Floquet
map
\begin{equation}
\delta x(nT)=F^{n}\,\delta x(0)\,,
\end{equation}
with the Floquet matrix F determining the evolution of the initial
set of variables $x(0)=(Re\delta c_{n}(0), Im \delta
c_{n}(0), \delta r_{n}(0),\delta
\dot{r}_{n}(0), \delta \theta_{n}(0), \delta
\dot{\theta}_{n}(0))$ after times $nT$. To ensure linear
stability all eigenvalues $\exp(i\,\psi_n)$ of the Floquet
matrix $F$ have to lie on the unit circle, i.e. the Floquet arguments
$\psi_n$ have to be real-valued. Intensive
numerical investigations in a wide parameter range have proved
that all the polarons gained from the map method are
linearly stable.

Moreover the Floquet analysis supplies the frequencies of the
normal modes via the relation
$\omega_{\psi}=\omega[\psi/(2\pi)+m]$ with $\psi$ expressed
in $\it{rad}$ and $m$ is an arbitrary integer number. We
identified the pinning modes for given sets of parameters.
According to \cite{Chen} we initiate the motion of polarons
through suitable perturbations of the velocity variables
$\dot{r}_{n}(t)$ and/or $\dot{\theta}_n$ targeted in the direction
of the pinning mode, that is we use for the numerical integration
of the system (\ref{eq:dotc})-(\ref{eq:dottheta}) the following
initial conditions
\begin{equation}
\left\{\,\Phi_{n},0,r_{n}^{(0)},0, \theta_{n}^{(0)},0\,\right\}+
 \left\{0,0,0,\lambda_r \xi_r,0,\lambda_{\theta} \xi_{\theta}\right\}\,,
\end{equation}
with the normalized momentum parts $\xi_{r,\theta}$  of the
pinning mode and $\lambda_{r,\theta}$ are the perturbation strengths (for
details see \cite{Chen}).

\noindent \underline{Ordered case}

In Fig.~\ref{fig:fig3}
 we present results for the
activation of breather motion along DNA in the ordered case. We
integrated the set of coupled nonlinear equations
(\ref{eq:dotc})-(\ref{eq:dottheta}) with a fourth-order
Runge-Kutta method and the accuracy of the computation was checked
through monitoring the conservation of the total energy as well as
the norm $\sum_{n}\,|c_n(t)|^2=1$. The applied kicking strength is
$\lambda_{r}=0.02$ corresponding to an amount of radial kinetic
energy of the order of $200\,meV$. Interestingly, we found that
there exists no  angular pinning mode component $\xi_\theta$  (see
further below) so that alone the excitation of the radial
component $\xi_{r}$ produces long-lived stable polaron and
breather motions. Hence, our findings suggest that higher
frequency radial fluctuational modes of DNA instigate the
propagation of localized structures along DNA rather than low
frequency twist fluctuational modes. The Fig.~\ref{fig:fig3}~(a)
  demonstrates that the electronic component of the
polaron moves with uniform velocity along the lattice maintaining
its localized shape throughout the journey. In the same manner the
radial polaron part travels as a localized wave along the strand
so that the electron propagation is accompanied by a local and
temporal contraction of the radii. However, the angular static
polaron component splits up into two parts when the radial pinning
mode is initially imposed. At the starting position there remains
a standing  breather of comparatively large amplitudes which
reverse periodically their sign in the course of time. Thus the
helix experiences alternately a local winding and unwinding around
its central base pair. In addition a second breather of relatively
small and exclusively negative amplitude propagates in unison with
the electronic and radial polaron components. The result is that
 the region of
the helix which is traversed by the electron gets locally
untwisted. As the quantitative assessment of the charge transport
is concerned we note that the breather travels across seventy
sites (base pairs) in a thousand (dimensionless) time units
meaning that physically the charge conduction across the base
pairs of DNA over a distance of $238\AA$ takes $\sim \, 0.16 ns$
leading to a higher transport rate than a typical one found for
protein conductivity \cite{Ye}. This underlines the status
ascribed to DNA molecules as promising candidates establishing
fast and efficient charge transport in (albeit only theorized yet)
molecular electronical engineering utilizing them as molecular
wires.

\noindent \underline{Static diagonal disorder}

In the presence of diagonal static disorder, that is when the
on-site electronic  energies $E_n^0$ are represented by random
numbers distributed in the interval  $|E_n^0|<\Delta E$ with width
$\Delta E$, we find that mobile breathers exist up to a critical
degree of disorder $\Delta E_{crit}$ and conductivity
persists. (Let us stress that it was indeed experimentally found
that the random sequence $\lambda-$DNA is an electrical
conductor \cite{Fink}.) Beyond this critical value continuous
propagation of the breathers along the lattice is prevented.

\begin{figure}
\begin{center}
    \includegraphics[angle=90,width=\doublefig]{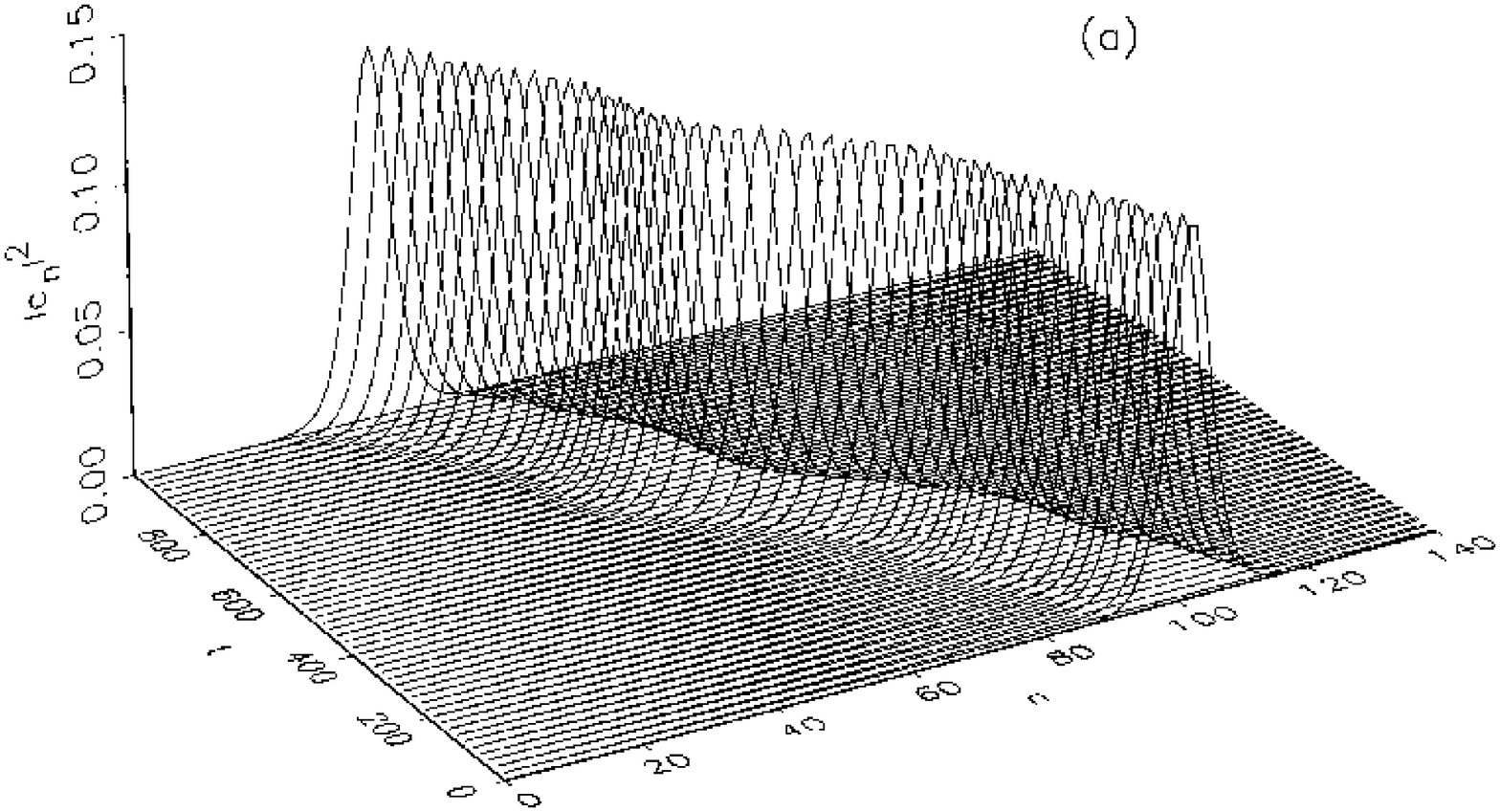}
    \mbox{}
    \includegraphics[angle=90,width=\doublefig]{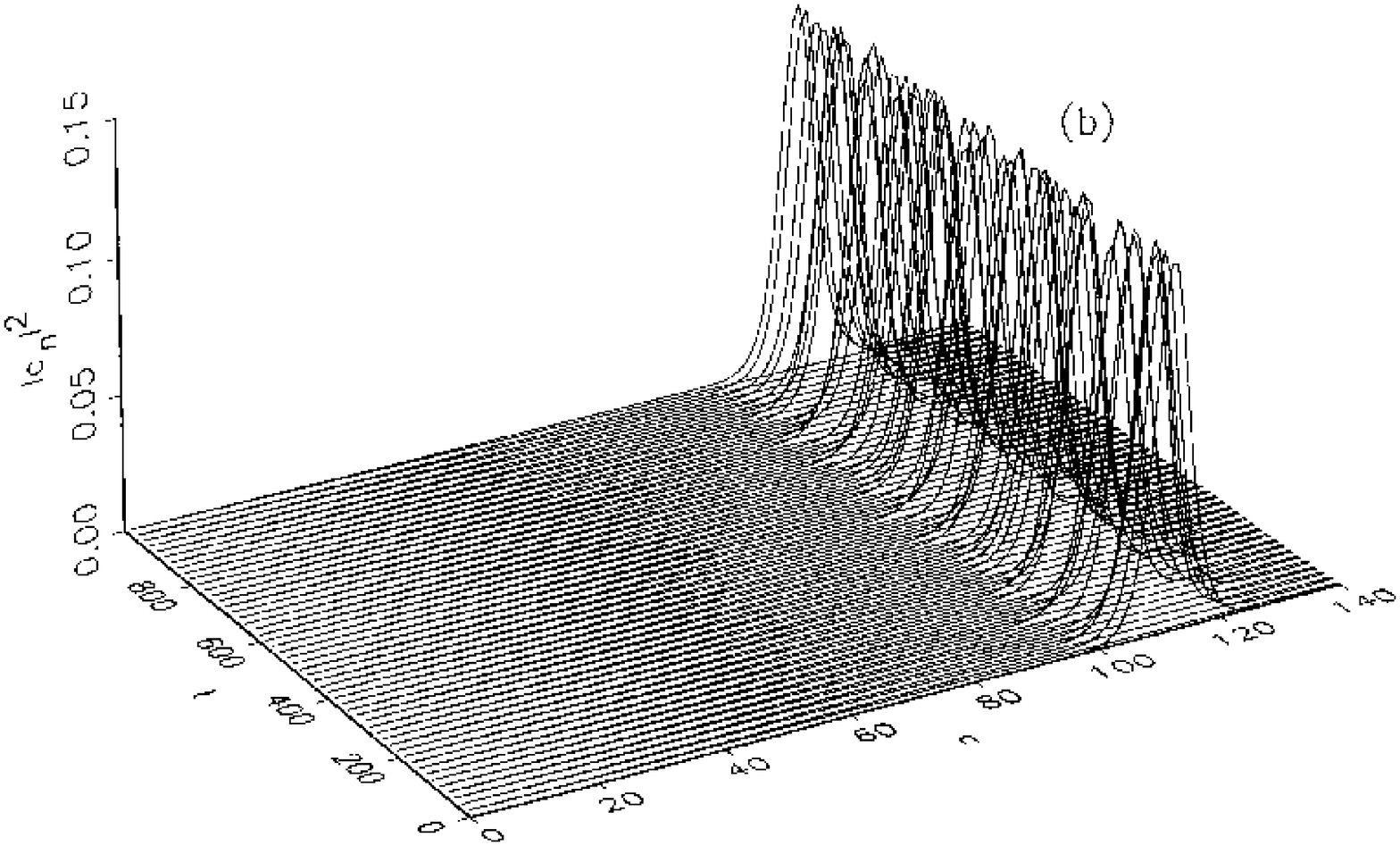}
\caption{Electron breather motion along DNA with diagonal disorder
for two different values of the width $\Delta E$ of the random
distribution of the on-site electronic energy $|E_n^0|$.
\newline (a) Width $\Delta E=0.05$. The electron breather moves
with uniform velocity along a strand.
\newline (b) Width $\Delta E=0.5$.  Confined electron breather
with positions oscillating itinerantly around the initial lattice
position. }
  \label{fig:fig4}
\end{center}
\end{figure}

This behavior is illustrated in Fig.~\ref{fig:fig4}
  displaying the
spatio-temporal evolution of the electronic amplitude $|c_n(t)|^2$.
While for a relatively
small amount of disorder $\Delta E=0.05$ the electronic breather
(together with its radial and torsional counterparts) is able to
travel uniformly along the lattice for a ten times increased
$\Delta E=0.5$ the
 breather motion is restricted to a
neighborhood of the initially excited position
around which it wanders  in an
oscillatory manner.

\begin{figure}[h]
\begin{center}
    \includegraphics[angle=90,width=\singlefig]{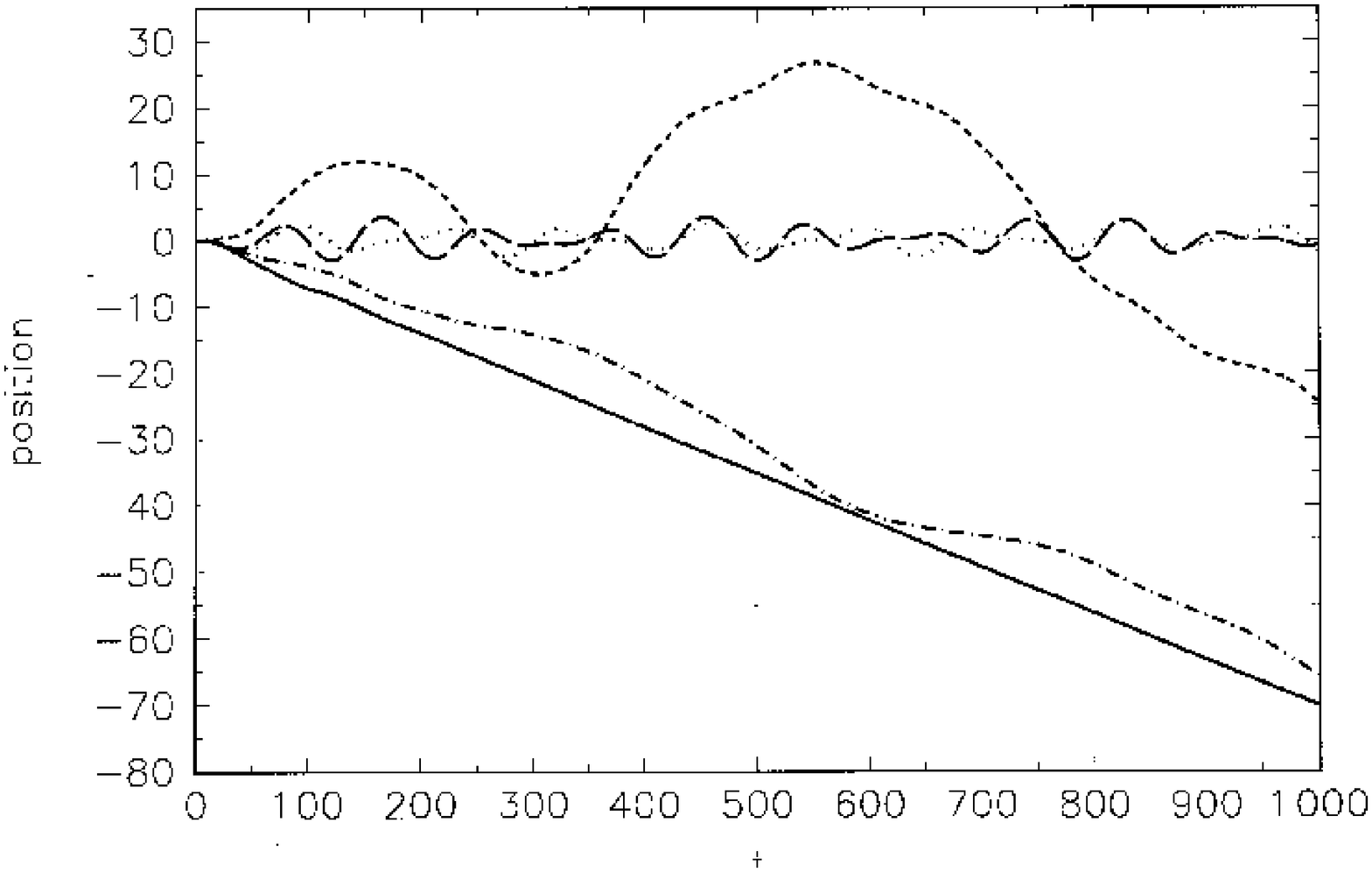}
\caption{Diagonal disorder. The position of the center of the
electron breather as a function of time and for different amounts
of disorder. Full line (Ordered case), short dashed line ($\Delta
E=0.025$), dashed-dotted line ($\Delta E=0.050$), dotted line
($\Delta E=0.100 $) and long dashed line ($\Delta E=0.500$). }
  \label{fig:fig5}
\end{center}
\end{figure}

Regarding mobility there act two competing mechanisms in the
combined disordered nonlinear system, namely a linear
Anderson-type one related with the effects of disorder which
attempts to pin the modes (bear in mind that the linear
Anderson-modes are supposed to be immobile) and opposed to this
there operates the nonlinear mechanism utilizing the pinning mode
to activate motion. The relative strength of these two mechanisms
$\Delta E/\lambda_{r}$ determines then breather mobility. In order
to summarize the results for the breather mobility in the presence
of diagonal disorder we display in \ref{fig:fig5}
  the temporal
behavior of the first momentum of the electronic occupation
probability defined as
\begin{equation}
\bar{n}(t)=\sum_{n}\,n |c_{n}(t)|^2\,,
\end{equation}

which determines the position of the breather center as a function
of time. The breathers are mobile and propagate unidirectionally
with (virtually) constant velocity along the lattice up to an
amount of disorder $\Delta E \simeq 0.045$. Generally, the larger
the width $\Delta E$ the lower becomes this velocity. Finally, for
an overcritical amount of disorder the breather performs
oscillatory motions around its starting lattice position. The
frequency of these oscillations diminishes gradually with more
enlarged $\Delta E$ and simultaneously the excursions from the
starting site get continuously stronger confined. Despite the
non-directed character of its motion even for a moderate amount of
disorder $\Delta E=0.05$ the breather is able to travel as far as
twenty lattice sites away from its starting position. Eventually
linear Anderson-type disorder effects
 overwhelm more and more the
nonlinear  activation mechanism for  breather motion through
the excitation of the nonlinear pinning mode.

\noindent \underline{Random double helix structure}

Concerning the impact of structural disorder we considered random
distributions of the base pair spacings $r_{n}-r_{0} \in [-\Delta
,\Delta ]$ and the twist angles $\theta_{n}-\theta_0 \in [-\Delta ,\Delta]$ with
different mean standard deviations $\Delta$.

\begin{figure}
\begin{center}
    \includegraphics[angle=90,width=\singlefig]{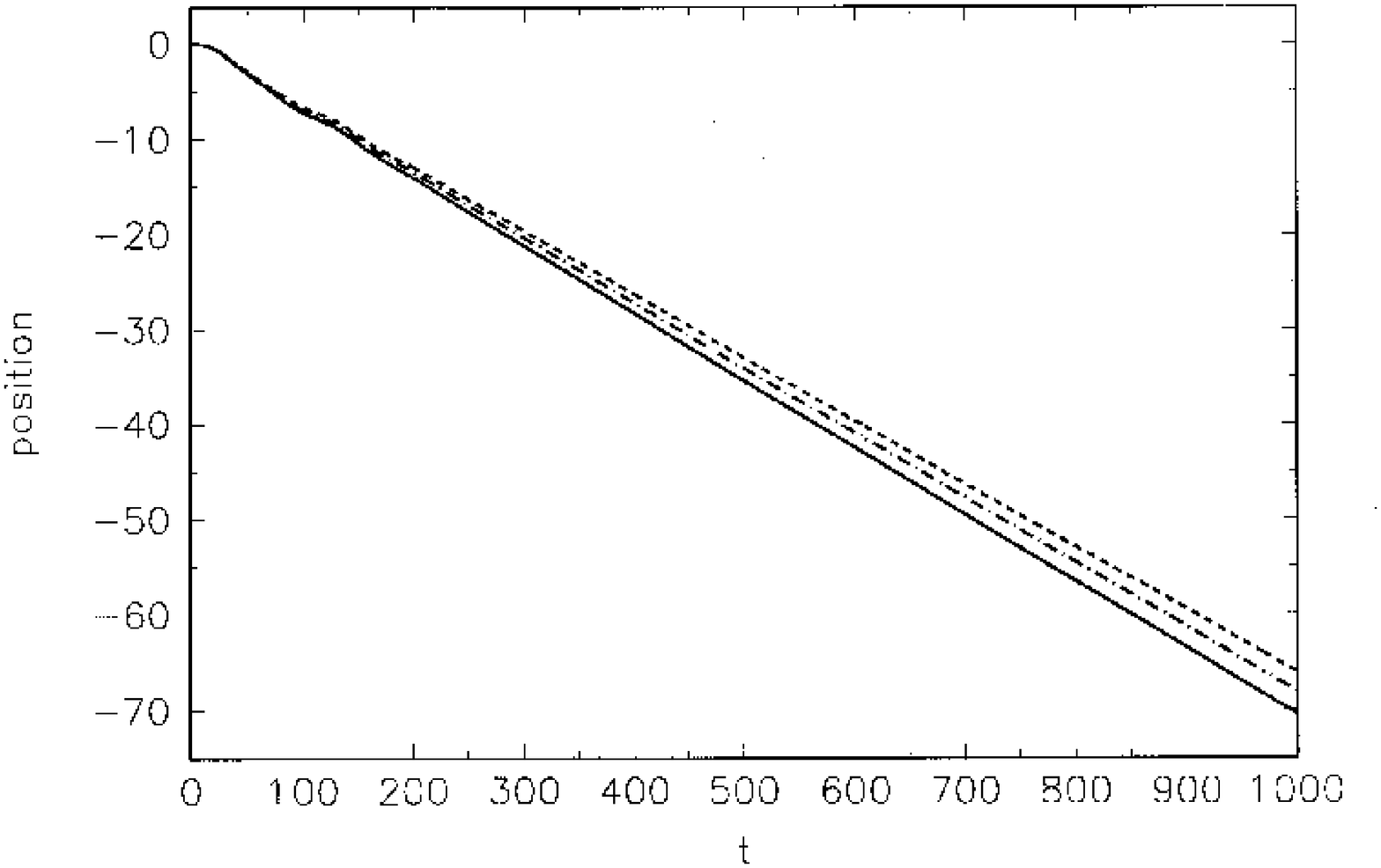}
\end{center}
\caption{Structural disorder and random double helix structure.
The temporal behavior of the position of the electron breather
center for different degrees of randomness in the radii $r_n$ and
angles $\theta_n$. Assignment of the line types to the increasing
mean standard deviations ranging up to $10\%$ of the respective
mean value: Full line ($1\%$), dashed dotted line ($5\%$) and
dashed line ($10\%$).}
  \label{fig:fig6}
\end{figure}

For representative results we averaged over several realizations
of structural disorder. The results for the temporal evolution of
the electron breather center are depicted in Fig.~\ref{fig:fig6}.
  We
conclude that structural disorder does not  significantly affect
the mobility of the breathers. For growing degree of disorder
merely the propagation velocity gets reduced. Even in the case
of fairly strong disorder $\Delta=10\%$ (dashed line) the breather
is still able to propagate steadily.

 We studied also the polaron mobility under the impact
 of combined static diagonal and structural
disorder and found that the strong pinning effect generated by static diagonal disorder
prevails over the more harmless and merely velocity-reducing influence of structural
irregularities.
Particularly the last result is
in compliance with the findings in \cite{Ye1} where it was reported that hopping
conductivity
in proteins and DNA is influenced first of all by the
molecular conformations and not by their sequences.

\section{Floquet analysis}

It is instructive to consider also the linear limit case of the  coupled nonlinear
electron-vibration system even though  the results
obtained throughout this paper bear essentially nonlinear character. The study of the linear
limit system itself  and its Floquet analysis can be applied as a reference and as a
tool for obtaining useful information for the interpretation of the Floquet analysis
of the fully nonlinear system.

\subsection{The linear system}\label{sub:linear}

The linear limit system is derived from the Eqns. (\ref{eq:dotc})-(\ref{eq:dottheta})
for $\alpha=k=0$ giving
\begin{eqnarray}
i \,\tau\,\dot{c}_{n}&=&-c_{n+1}-c_{n-1} \label{eq:dotlinealc} \\
\ddot{r}_{n}&=&-r_n \label{eq:dotlinealr}\\
\ddot{\theta}_{n\,n-1}&=&-\Omega^2\,\theta_{n\,n-1}\ \label{eq:dotlinealth}
\end{eqnarray}
In the first equation the terms $E_n^0\,c_n$, which in a
homogeneous system reduce to $E_0\,c_n$, have been eliminated through the
gauge transformation
$c_n \rightarrow c_n \exp (-i\,{E_0\,t}/{\tau})$.
In the linear limit
the electronic and vibrational degrees of freedom are decoupled.

To construct the electronic state we substitute in Eq.
(\ref{eq:dotlinealc}), $c_n(t)=\phi_n \exp (-iE\,t/\tau)$ with
time-independent $\phi_n$
 yielding  the linear stationary discrete Schr\"odinger equation:
 $E\,\phi_n=-\phi_{n+1}-\phi_{n-1}$\,.
With the Bloch mode solutions  $\phi_n=\exp(i\,q\,n)$ we get then
the dispersion relation: $E=-e^{i\,q}-e^{-i\,q}=-2\,\cos q$\,. The
linear electronic energy spectrum runs from $-2$ to $2$. The
polaron in the full system displayed in Fig.~\ref{fig:profile}
possesses an energy value  $E=-2.064$ which is slightly below the
lower band edge $E=-2$ assigned to  a wave number  $q=0$ and in
the linear regime the frequencies are in the range
$w_p={E}/{\tau}\in [-7.75,7.75]$. Conclusively,
 the time scale for the evolution of the angular (radial) variables $\theta_{n\,n-1}$
($r_n$) represented by
independent harmonic oscillators with frequencies $1$ and $\Omega=0.0842$,
respectively,
 is two orders (one order) of magnitude larger than the one of
the electron.

The linear analysis is also useful to understand the origin of many of
the numerically calculated Floquet arguments for the fully nonlinear
 system.  The
values of the variables $r_n$ and $\theta_{n\,n-1}$ after a period
$T=2\,\pi/w_p=2\,\pi\,\tau/E$ are:
$r_n(T)=\exp(i\,2\,\pi\,/w_p)\,r_n(0)$ and
$\theta_{n\,n-1}(T)=\exp(i\,2\,\pi\,\Omega/w_p)\,\theta_{n\,n-1}(0)$.
Therefore, their corresponding Floquet arguments are given by
$2\,\pi/w_p$ and $2\,\pi\,\Omega/w_p$, for $r_n$ and $\theta_{n\,n-1}$,
respectively, which are attributed to angular positions in the complex
plane with values around $\pm 45^0$ and $\pm 3.9^0$ for polarons with
energies slightly below $E=-2$ and with increased $|E|$ these angular
positions shift to lower values on the unit circle.

\subsection{Results of the Floquet analysis}
\label{sec:floquet}

The numerical integration of the system of tangent equations over one
period of the polaron's oscillation $T=2\,\pi\,\tau/|E|$ allows us to
obtain the actual Floquet matrix. Its eigenvalues can be written as
$\exp(i\,\psi_n)\,$ with $\psi_n$ being the Floquet arguments. The
system is linearly stable provided all the Floquet arguments are real.
As noted earlier in this paper within the range of parameters studied in
this paper linear stability is assured, but this is not the only
information that can be gained from the Floquet analysis.

In order to relate the positions of the numerous eigenvalues of the
Floquet matrix in the complex plane to the various degrees of freedom
contained in the system it is elucidating to study first an
anti-coupling limit in the electronic subsystem. Introducing an
auxiliary tunable coupling parameter $0\le \mu\le 1$, which corresponds
to the transfer integral in the scaled system, the anti-coupling limit
is obtained with $\mu=0$ and the actually studied coupled system with $\mu=1$.

\begin{figure}
  \begin{center}
    \includegraphics[width=\doublefig]{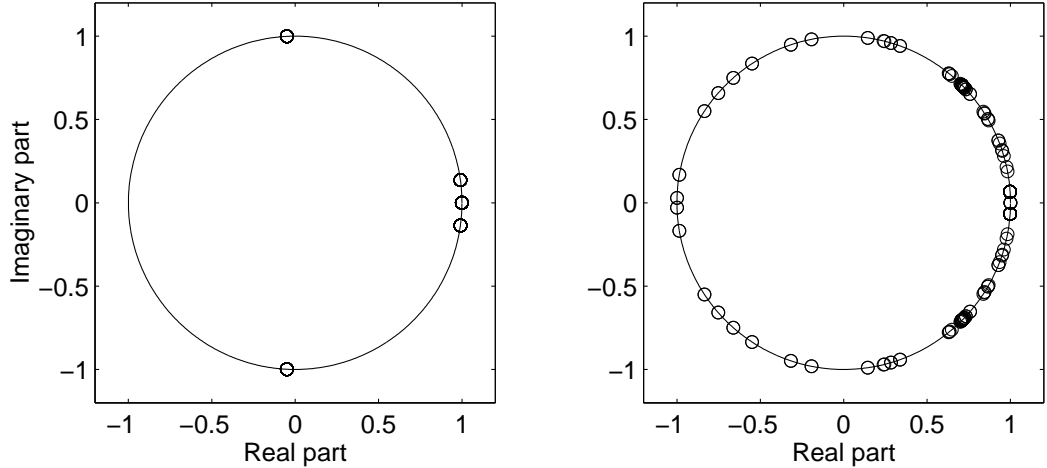}
  \end{center}
  \caption{The Floquet
multipliers in the complex plane. Left figure: The linear limit
system (\ref{eq:dotlinealc})-(\ref{eq:dotlinealth}) at the
anti-coupling limit $\mu=0$. Right figure: The fully coupled
nonlinear system.}
  \label{fig:floquet1}
\end{figure}

For the decoupled electron system ($\mu=0$) the value of
the energy $E$ can be  easily derived with the
help of the Eqs. (\ref{eq:rinst}) and (\ref{eq:statthree}), giving
$E\Phi_n=-k^2|\Phi|^2 \Phi_n$ and $r_n^{(0)}=-k|\Phi_n|^2$, which leads
 to $E=-1$ for the localized state $\Phi_n=\delta_{n\,0}$ and $k=1$ as a typical value.
Therefore, the $r_n$-multipliers are in the
vicinity of $\pi/2$, the $\theta_{n\,n-1}$'s lie at small angles at both
sides of $(1,0)$
 and the $c_n$'s are located at $(1,0)$ on the unit circle, in
accordance with the analysis performed in \ref{sub:linear}. As we
switch on the coupling $\mu$ and increase it up to the actual
value $\mu=1$, we can follow the evolution of the multipliers.
Eventually, for $\mu=1$ the $r_n$'s and $\theta_{n\,n-1}$'s have
moved to positions with about half their initial angle values
while spreading a little  on the unit circle (the
$\theta_{n\,n-1}$'s to lesser extent though). The $c_n$'s separate
from $(1,0)$ and invade the whole circle, as a consequence of the
nonlinear coupling between the electron amplitudes and the
vibrations. The exceptions are the pair at $(1,0)$, corresponding
to the phase mode, and another isolated pair close to this point.
The latter pair is associated with the spatially antisymmetric
pinning mode, responsible for the mobility of the polaron. Note
that the linear approximation described in \ref{sub:linear} is an
approximation for the radial and angular variables. The Floquet
multipliers for the decoupled and full systems, respectively are
shown in Fig.~\ref{fig:floquet1}.

\begin{figure}
  \begin{center}
    \includegraphics[width=\doublefig]{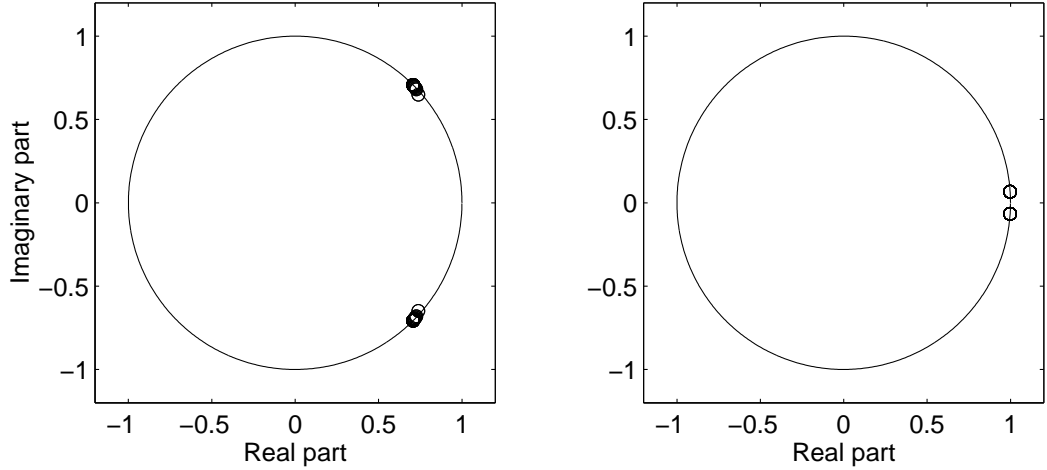}
  \end{center}
  \caption{Multipliers of $F_{r\,r}$ (left figure) and
$F_{\theta\,\theta}$ (right figure).}
  \label{fig:floquet2}
\end{figure}

\begin{figure}
  \begin{center}
    \includegraphics[width=\doublefig]{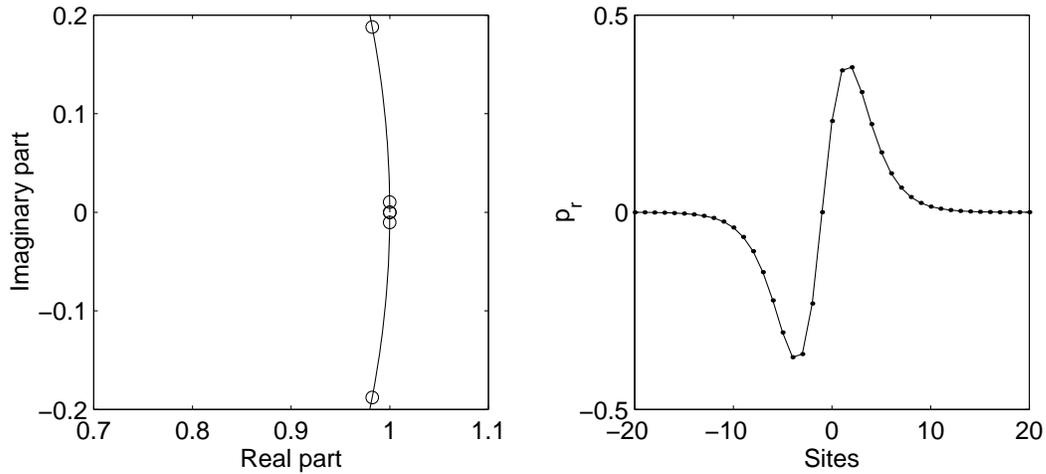}
  \end{center}
  \caption{Multipliers of  $F_{c\,r}$  in the vicinity of $(1,0)$.
The pinning mode eigenvalues are recognizable. The right part of
the figure shows the normalized pinning mode.}
  \label{fig:floquet3}
\end{figure}

In order to illuminate further the role played by the vibrational modes
 with respect to
the initiation of mobility of the polaron we
cast the Floquet matrix $F$  in the form of a  supermatrix formed by submatrices:
\begin{equation}
F= \left(\begin{array}{cccc}
    F_{c\,c} & F_{c\,r} & F_{c\,\theta}\\
    F_{r\,c} & F_{r\,r} & F_{r\,\theta}\\
    F_{\theta\, c} & F_{\theta\, r} & F_{\theta\,\theta}\\
    \end{array} \right)\,,
\end{equation}
and the indices of the submatrices refer to the origin of their
entries. From the arguments of the eigenvalues of the submatrices
an/or combinations of them, it is possible to deduce additional
information about the spectral features attributed to the various
degrees of freedom contained in the coupled nonlinear system. The
matrix $F_{c\,c}$ appears in the Born-Oppenheimer approximation.
To recover the pinning mode of the complete system it suffices to
consider only the subsystem consisting of the electronic part
coupled to the radial motions for which the Floquet matrix reduces
to the submatrix $F_{C\,R}=(F_{c\,c}\,\, F_{c\,r}; F_{r\,c}
\,\,F_{r\,r}) $, which means that the mobility of the polaron is
mainly a consequence of the coupling between the electron
amplitudes and the radial variables and that the electron will not
move within the Born-Oppenheimer approximation. The
Figs.~\ref{fig:floquet2} and \ref{fig:floquet3}

illustrate these statements. The left (right) part of the
Fig.~\ref{fig:floquet2} shows the multipliers of $F_{r\,r}$
($F_{\theta\,\theta}$). In comparison with the linear case studied
in the preceding section we note that while the radial components
of the multipliers shifted their positions to lower angles and
exhibit some spread the angular multipliers hardly alter their
position. Hence, perturbations of the angular twist motions in the
context of the
 coupled
system resemble the behavior of the linear limit system. The  left
part of Fig.~\ref{fig:floquet3} depicts the eigenvalues of
$F_{c\,r}$ in the vicinity of $(1,0)$. The pair of pinning mode
eigenvalues is recognizable. In the right part of the figure the
normalized pinning mode is portrayed.

Moreover, it should be pointed out that for
larger elongations of $\theta_{n\,n-1}$, beyond the range where the linear
expansion of $d_{n\,n-1}$ is valid, there exists also an
angular pinning mode as a
consequence of the coupling between the electronic amplitudes
and the angular variables.

\begin{figure}
  \begin{center}
    \includegraphics[width=\singlefig]{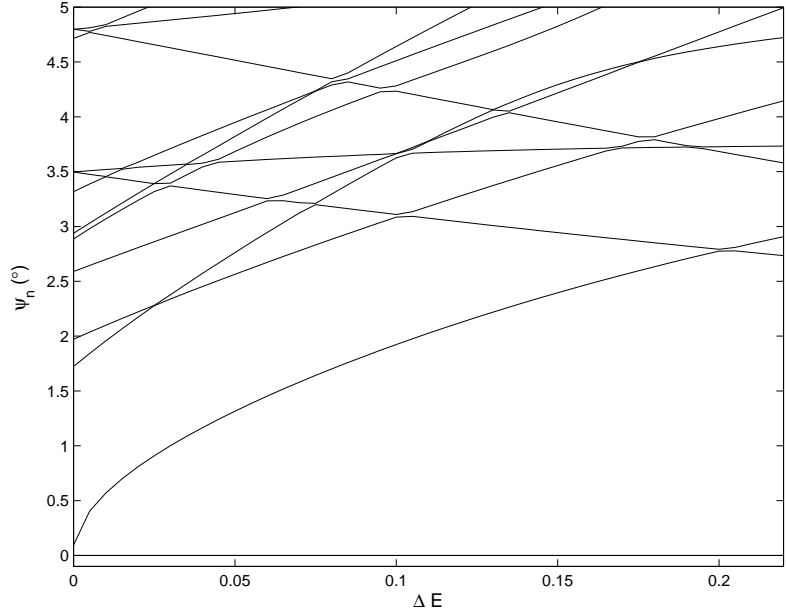}
  \end{center}
  \caption{Arguments of the Floquet multipliers of the low-frequency
modes as a function of diagonal disorder $\Delta E$ demonstrating
the growth of the argument of the pinning mode with increased
amount of disorder so that the pinning mode eventually merges in
the rest of the modes.}
  \label{fig:fdisorder}
\end{figure}

The fact that the disorder eventually prevents the polaron from
being moved can also be explained with the help  of the Floquet
multipliers. When the degree of disorder is increased the pinning
mode eigenvalues move progressively away from $(1,0)$ towards
larger arguments $\psi$ and approach the rest of the $F_r$
eigenvalues. Fig.~\ref{fig:fdisorder}
 displays the growth of the
arguments of the pinning mode as a function of the
disorder strength $\Delta E$. Eventually the system is too far away from the
instability due to the marginal mode, which appears when the pair of
pinning mode eigenvalues collapses at $(1,0)$. This is easy to
understand, as the disorder brings about a decoupling of the variables,
favoring localized, stationary, isolated modes.

\section{Summary}\label{section:summary}
\enlargethispage{1cm}
In the present work we have studied an electron transfer mechanism in
DNA relying on the coupling of the charge carrying unit to the
vibrational modes of DNA responsible for the formation of polarons and
electron-vibron breathers. The steric structure of the double helix of
$\lambda-$DNA has been described in the context of the base-pair picture
considering angular and radial vibrational motions of the base pairs. We
have focused interest on the initiation of long-range and stable polaron
and breather motion along the DNA structure when static and/or
structural disorder inherent to any real DNA molecule is taken into
account. We remark that the proposed polaron-like transport mechanism
applies to electron hole conduction too. In the first part of the paper
we have constructed exact stationary localized solutions (polarons and
static electron-vibron breathers) with the help of a nonlinear map
approach. We have found that the interplay of disorder and nonlinearity
amplifies the degree of localization in comparison with cases when
either disorder or nonlinearity alone is present. With concern to mobile
electron-vibron breather solutions, whose motion is activated through a
kick mechanism utilizing the pinning mode, with the current work we have
demonstrated that they are fairly robust when the electron-vibration
system is subjected to randomness. Generally, even for a moderate amount
of disorder the breathers retain their localized shapes and support
directed, long-ranged coherent electron transfer along strands of the
bent double helix. In particular the last result suggests that DNA seems
suitable for the design of (one-dimensional) functional nanostructures
as ingredients of molecular nanoelectronical devices. Furthermore, we
have observed that the electron breather is accompanied by vibrational
breathers in the radial as well as angular components, respectively
causing local and temporal deformations of the traversed region of the
double helix. Conclusively,
 efficient charge transport in DNA proceeds on the condition that the
 double helix undergoes structural changes which is exemplary for the
interplay of structure and function in flexible and adaptive  biomolecules.

\vspace{0.5cm}
\centerline{\large{\bf Acknowledgments}}

\noindent One of the authors (D.H.) acknowledges  support by the
Deutsche Forschungsgemeinschaft via a Heisenberg fellowship
 (He 3049/1-1). J.F.R.A. acknowledges D.H. and the
 Institut f\"{u}r Theoretische Physik for their warm hospitality.
J.A. is grateful for a scholarship supplied by the
 Studienstiftung des deutschen Volkes.
 The authors are also
grateful to the support under the LOCNET EU network
HPRN-CT-1999-00163.


\end{document}